\documentclass[twoside,fleqn]{article}
\usepackage{espcrc2}
\usepackage{epsfig}

\newenvironment{CFigure}[1][h]{\begin{figure}[#1]\centering}%
                                {\end{figure}}

\def\gsim{{\mathrel{\raise2pt\hbox to 8pt{\raise -5pt\hbox{$\sim$}\hss{$>$}}}}}
\def\rsim{{\mathrel{\raise2pt\hbox to 8pt{\raise -5pt\hbox{$\sim$}\hss{$>$}}}}}
\def\lsim{{\mathrel{\raise2pt\hbox to 8pt{\raise -5pt\hbox{$\sim$}\hss{$<$}}}}}

\def\etal{{\it et al.}}

\begin{document}

\title{
       \begin{flushright}\normalsize
	    \vskip -0.9 cm
            UW-PT 03-12
       \end{flushright}
	\vskip -0.4 cm
Extending the Utility of Partially Quenched QCD
\thanks{Presented by S.~Sharpe.
Research supported in part by US-DOE contract DE-FG03-96ER40956/A006.}
}

\author{Stephen R. Sharpe$\rm ^{a}$ and Ruth S. Van de Water\address{Physics Department, Box 351560,
University of Washington, Seattle, WA 98195-1560, USA}}
      
\begin{abstract}
It has been proposed that partially quenched chiral perturbation theory can be used together with partially quenched lattice QCD to determine the low-energy constants of \emph{real} QCD.  We point out a complication of this approach and show how it can be resolved.  Partial quenching changes the chiral symmetry group and introduces a new independent four-derivative operator.  We explore the effects of this new operator. We describe tree-level scattering processes to which it contributes, and show how they can be used, in principle, to deterimine its unknown coefficient.  We also calculate its lowest-order (one-loop) contribution to charged meson masses and decay constants.  In addition, we present the form of the analytic next-to-next-to-leading order meson mass and decay constant corrections, which are needed for lattice fits.
\vspace{-0.2in}
\end{abstract}

\maketitle

\section*{Introduction and Overview}

Chiral perturbation theory ($\chi$PT) allows analytic calculation of low-energy QCD processes involving the light pseudoscalar mesons.  Results are given in terms of a number of undetermined coefficients, such as the Gasser-Leutwyler coefficients and $f_\pi$, which we would like to determine from first principles.  Lattice QCD provides a method for doing this as long as simulations are done at low enough quark masses that $\chi$PT (typically at next-to-leading order) is a good approximation.  In practice, however, lattice simulations are not done at the actual QCD values of the quarks masses because simulating light, dynamical quarks is computationally expensive.  Thus there is a limited range of quark masses where both lattice simulations are feasible and $\chi$PT describes the physics.  

This situation can be improved using the partially quenched (PQ) approximation, in which valence quarks (those which appear in external states) and sea quarks (those which appear in dynamical loops) are allowed to have different masses.  Allowing valence and sea quarks to have different masses expands the parameter space available for determining the QCD low-energy constants, as shown in Fig.~\ref{fig:Plot}.     
\begin{CFigure}
	\epsfysize=2.3in
	\vspace{-0in}\hspace{-.0in}\rotatebox{0}{\leavevmode\epsffile{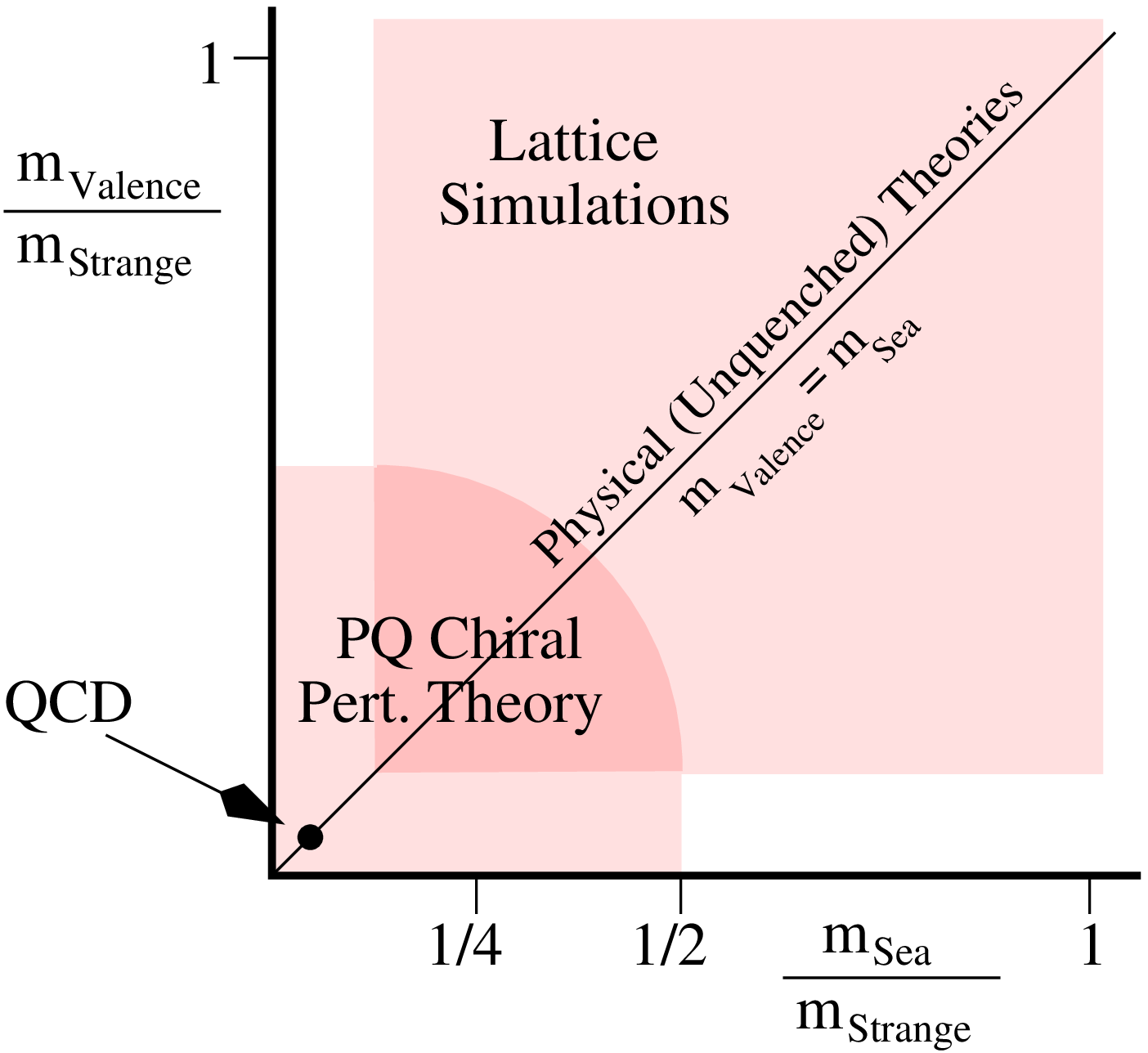}}
	\vspace{-.4in}\caption{Partially quenched parameter space}\vspace{-.25in}
	\label{fig:Plot}
\end{CFigure}

It has been argued in Refs.~\cite{PhysicalResults,Phi0} that although PQ QCD is unphysical, the properties of pseudo-Goldstone bosons (PGBs) can be expressed in terms of the physical coefficients of the QCD chiral Lagrangian through next-to-leading order (NLO) in $\chi$PT.  While the conclusions about NLO masses and decay constants are correct, the argument is incomplete.  We point out an overlooked subtlety in using PQ simulations to extract physical QCD constants, and discuss the consequences.

\vspace{-0pt}\section*{New Four-Derivative Operator in PQ$\chi$PT}

Partial quenching introduces unphysical bosonic ghost quarks and therefore changes the low-energy symmetry structure
\cite{Bernard}.  The chiral symmetry group becomes a graded group containing both commutation and anti-commutation relations:
\begin{equation}
	SU(N_{Sea}) \rightarrow SU(N_{Valence} + N_{Sea}| N_{Valence})
\end{equation}
We observe that, because of the graded group structure, the four-derivative operator:
\begin{equation}
	\mbox{Str}(\partial_\mu U \partial_\nu U^{\dagger} \partial^{\mu} U \partial^{\nu} U^{\dagger})
\end{equation}
is linearly-independent, while in $SU(N)$, $N\leq3$, it is not.  For comparison with unquenched QCD, it is convenient to use a linear combination of operators that vanish in the unquenched $SU(3)$ sector of the PQ theory, with an undetermined coefficient, $L_{PQ}$:
\vspace{-1pt}
\begin{eqnarray}
	\mathcal{O}_{PQ} & \! = \! & L_{PQ} \{ \mbox{Str}(\partial_\mu U \partial_\nu U^{\dagger} \partial^{\mu} U \partial^{\nu} 	U^{\dagger}) \nonumber \\
	&& -\frac{1}{2} \mbox{Str}(\partial_\mu U \partial^\mu U^{\dagger})^2 \nonumber \\
	&& - \mbox{Str}(\partial_\mu U \partial_\nu U^{\dagger})\cdot \mbox{Str}(\partial^{\mu} U \partial^{\nu} U^{\dagger}) \nonumber \\
	&& + 2 \mbox{Str}(\partial_\mu U \partial^\mu U^{\dagger} \partial_\nu U \partial^\nu U^{\dagger}) \}
\label{eqn:Op}
\end{eqnarray}
In summary, we find that this new operator: 
\begin{itemize}

\item{contributes at NLO to ``unphysical'' scattering processes that cannot occur in unquenched $SU(3)$.}

\item{contributes at next-to-next-to leading order (NNLO) to PGB masses and decay constants.}

\end{itemize}

Because $\cal{O}_{PQ}$ vanishes in the unquenched $SU(3)$ limit, it cannot contribute at tree-level to ``physical'' scattering processes that can occur in the unquenched $SU(3)$ sector.\footnote{Here we do not consider processes involving neutral mesons because double poles in the meson propagators make it unclear how to amputate diagrams and thus define scattering amplitudes.}  For example, the process shown in Fig.~\ref{fig:ABScatt} is an $SU(3)$ process because it involves only two quark flavors.  The various quark line diagrams cancel each other out because of relative minus signs and numbers of Wick contractions.  However, $\cal{O}_{PQ}$ does contribute to the scattering process in Fig.~\ref{fig:4QuarkScatt} because the process involves four quark flavors.  In the case of PQ QCD with $N_{Sea}=3$, these could be four valence quarks, three sea quarks and a valence quark, or other possibilities. Figure~\ref{fig:GhostScatt} also receives contributions from $\cal{O}_{PQ}$ because this process is unique to the PQ theory, involving both valence and ghost quarks.  These examples show that, physically, the new particle content of the PQ theory allows separation of new combinations of quark contractions, necessitating an additional operator in the Lagrangian.

\begin{CFigure}
	\epsfysize=1.8in
	\vspace{-.27in}\hspace{-.0in}\rotatebox{0}{\leavevmode\epsffile{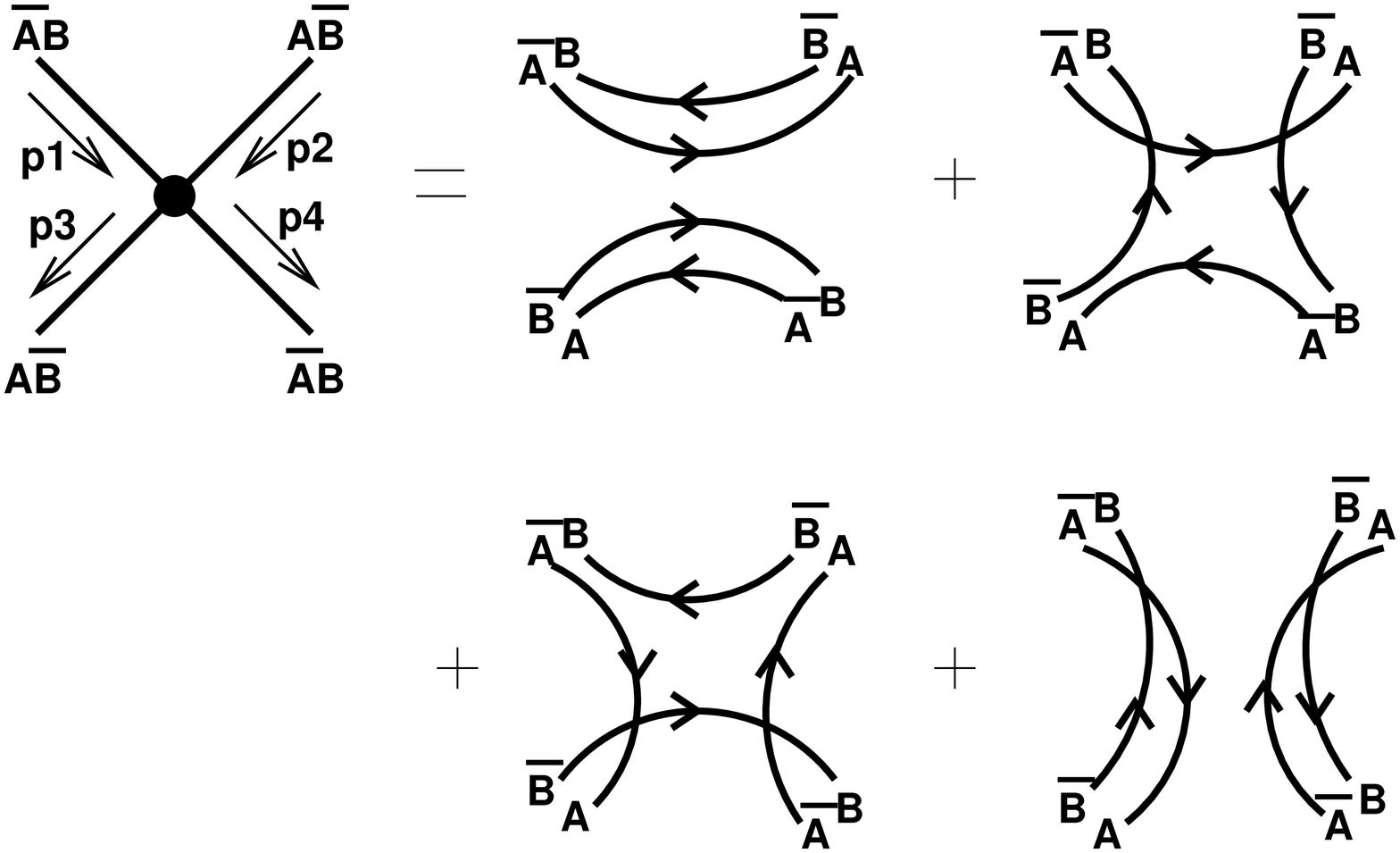}}
	\vspace{-.55in}\caption{``Physical'' scattering process because it involves only two quarks}\vspace{-.25in}
	\label{fig:ABScatt}
\end{CFigure}
\begin{CFigure}
	\epsfysize=.8in
	\vspace{-.34in}\hspace{-.0in}\rotatebox{0}{\leavevmode\epsffile{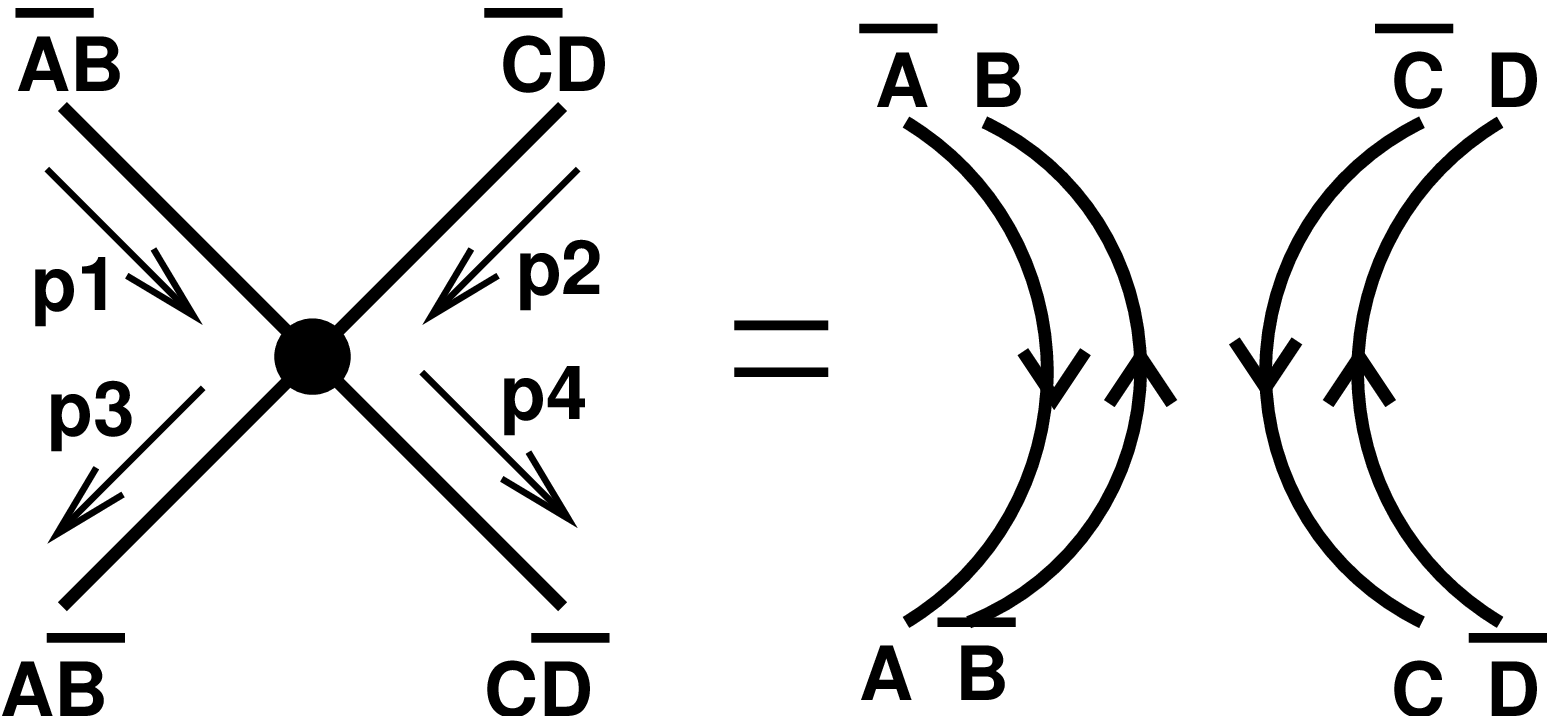}}
	\vspace{-.35in}\caption{``Unphysical'' scattering process because it involves more than three quarks}\vspace{-.3in}
	\label{fig:4QuarkScatt}
\end{CFigure}
\begin{CFigure}
	\epsfysize=.8in
	\vspace{-.32in}\hspace{-.0in}\rotatebox{0}{\leavevmode\epsffile{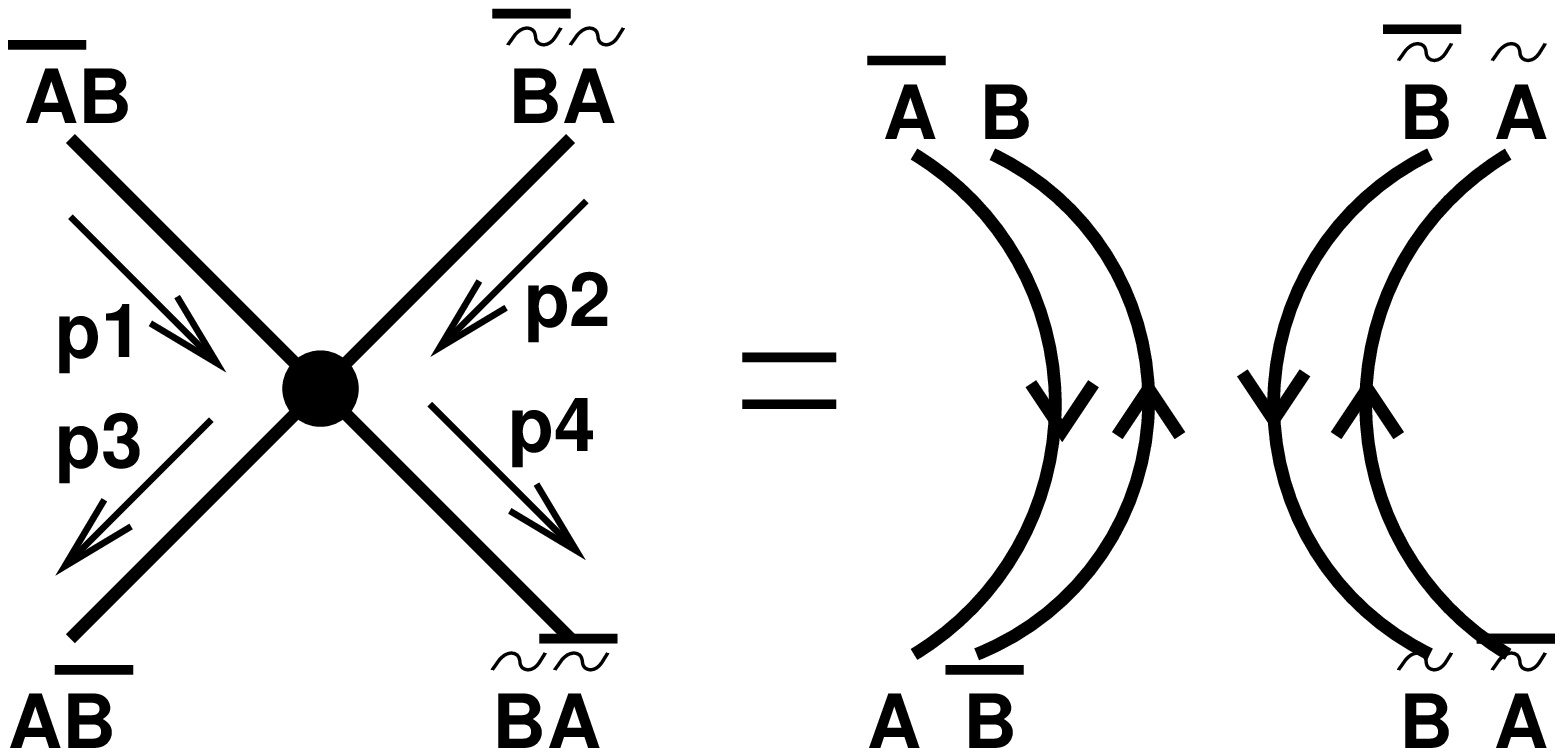}}
	\vspace{-.35in}\caption{``Unphysical'' scattering process because it involves both valence and ghost quarks}\vspace{-.25in}
	\label{fig:GhostScatt}
\end{CFigure}
Like $f_\pi$ and the Gasser-Leutwyler coefficients, $L_{PQ}$ can be obtained by fitting to lattice experiments.  To illustrate this we calculate meson scattering processes that only receive contributions from double supertrace operators in PQ$\chi$PT, such as that shown in Figure~\ref{fig:GhostScatt}.  The $\mathcal{O}(p^4)$ part of the scattering amplitude is: 
\begin{eqnarray}
%
\lefteqn{f^4 \mathcal{M}   =  
 32 L_1(p_1\cdot p_3)(p_2\cdot p_4)} \nonumber \\
	&& +16 L_2 \Big[(p_1\cdot p_2)(p_3\cdot p_4)+
			(p_1\cdot p_4)(p_2\cdot p_3)  \Big] \nonumber \\
	&& +16 L_{PQ} \Big[(p_1\cdot p_2)(p_3\cdot p_4)+
			   (p_1\cdot p_3)(p_2\cdot p_4)     \nonumber \\
	&& +(p_1\cdot p_4)(p_2\cdot p_3) \Big]
\end{eqnarray}
$L_1$, $L_2$, and $L_{PQ}$ can, in principle, be separately determined by combining this with other amplitudes and fitting to lattice data\footnote{Problems due to the lack of unitarity in the PQ theory \cite{Lin} do not appear until one-loop level, i.e. NNLO in $\chi$PT.}.

Because $\cal{O}_{PQ}$ is a four-derivative operator, it does not contribute to masses and decay constants until NNLO order in PQ$\chi$PT, and therefore does not affect the conclusions of Refs.~\cite{PhysicalResults,Phi0}  As a first step in extending the program of Refs.~\cite{PhysicalResults,Phi0} to NNLO, we calculate the lowest-order contribution of $\cal{O}_{PQ}$ to charged meson masses and decay constants\cite{Me}.  To simplify calculations we consider only two valence quarks, $A$ and $B$, and $N$ sea quarks of equal mass.  The charged meson, $\pi_{AB}$, receives NNLO corrections from the new operator to its mass from Figure~\ref{fig:Mass} and to its decay constant from Figure~\ref{fig:Decay}.  All of the meson diagrams receive contributions from the classes of quark line diagrams shown in Figure~\ref{fig:QL}, which illustrate the flow of quarks within the mesons.  Quark line diagrams show the effect of partial quenching | valence quarks appear only as external states because ghost quarks cancel them in loops.  Quark ``hairpins'' remove the flavor singlet propagator to enforce the fact that $SU(N|M)$ generators are traceless.  Both the mass and decay constant corrections from the new operator vanish in the unquenched $SU(3)$ limit ($m_A \! = \! m_B \! = \! m_{Sea} , \; N_{Sea} = 3$) as they should.  The expressions will appear in Ref.~\cite{Me}.  

\section*{Analytic NNLO Mass and Decay Constant Corrections}

Fits of present PQ lattice data require NNLO terms \cite{Farchioni1,Farchioni2}.  Full non-analytic NNLO calculations in PQ$\chi$PT are not available;  however, as an intermediate step we have determined the form of the analytic NNLO terms.  Although over a dozen operators in $\mathcal{L}_6$ contribute, only four linear combinations of quark masses appear in the tree-level correction to the mass of $\pi_{AB}$:
\begin{eqnarray}
	\Bigg( \!\frac{\delta m_{AB}^2}{m_{AB}^2} \!\Bigg) \!\!\!\!\! & = & \!\!\! \alpha_1 \chi_S^2 + \alpha_2 \chi_S \Big( 	\chi_A + \chi_B \Big) \nonumber \\
	& + & \!\!\! \alpha_3 \Big( \chi_A + \chi_B \Big)^2 + \alpha_4 \Big( \chi_A - \chi_B \Big)^2
\end{eqnarray}
Here $\alpha_i$ are linear combinations of $\mathcal{L}_6$ coefficients.  Corrections to $f_{AB}$ have the same form with different coefficients.  Fits to PQ lattice data with $\frac{1}{3}m_S < m_{Sea} < \frac{2}{3}m_S$ are consistent with this NNLO formula \cite{Farchioni2}.
\begin{CFigure}
	\epsfysize=0.7in
	\vspace{-.205in}\hspace{-.0in}\rotatebox{0}{\leavevmode\epsffile{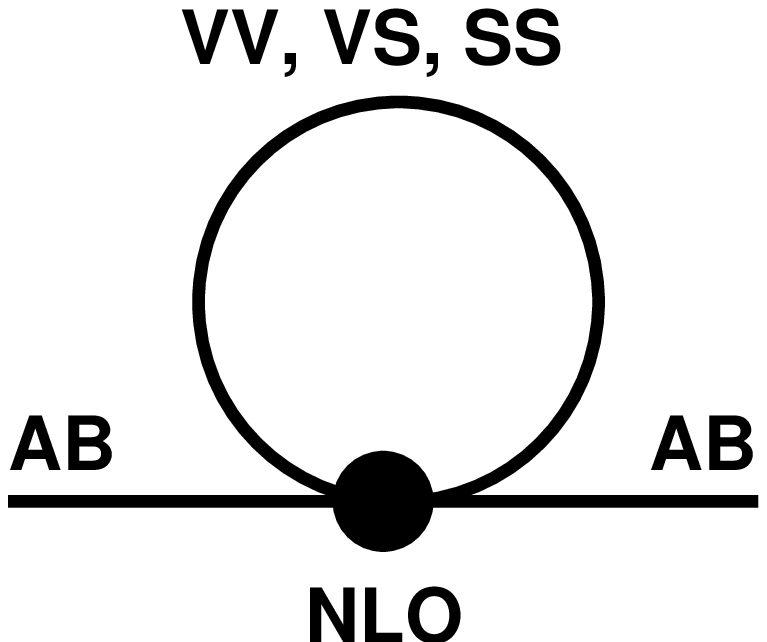}}
	\vspace{-.36in}\caption{PGB mass renormalization from $\cal{O}_{PQ}$}\vspace{-.4in}
	\label{fig:Mass}
\end{CFigure}
\begin{CFigure}
	\epsfysize=0.65in
	\vspace{-.165in}\hspace{-.0in}\rotatebox{0}{\leavevmode\epsffile{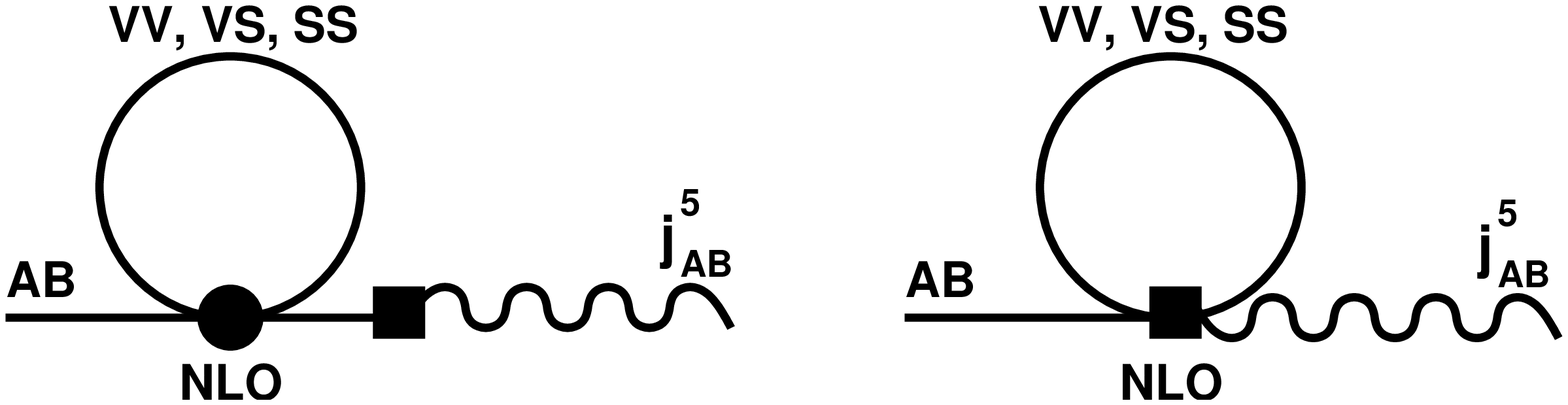}}
	\vspace{-.35in}\caption{PGB decay constant renormalization from $\cal{O}_{PQ}$}\vspace{-.22in}
\label{fig:Decay}
\end{CFigure}
\begin{CFigure}
	\epsfysize=1.9in
	\vspace{-.40in}\hspace{-.0in}\rotatebox{0}{\leavevmode\epsffile{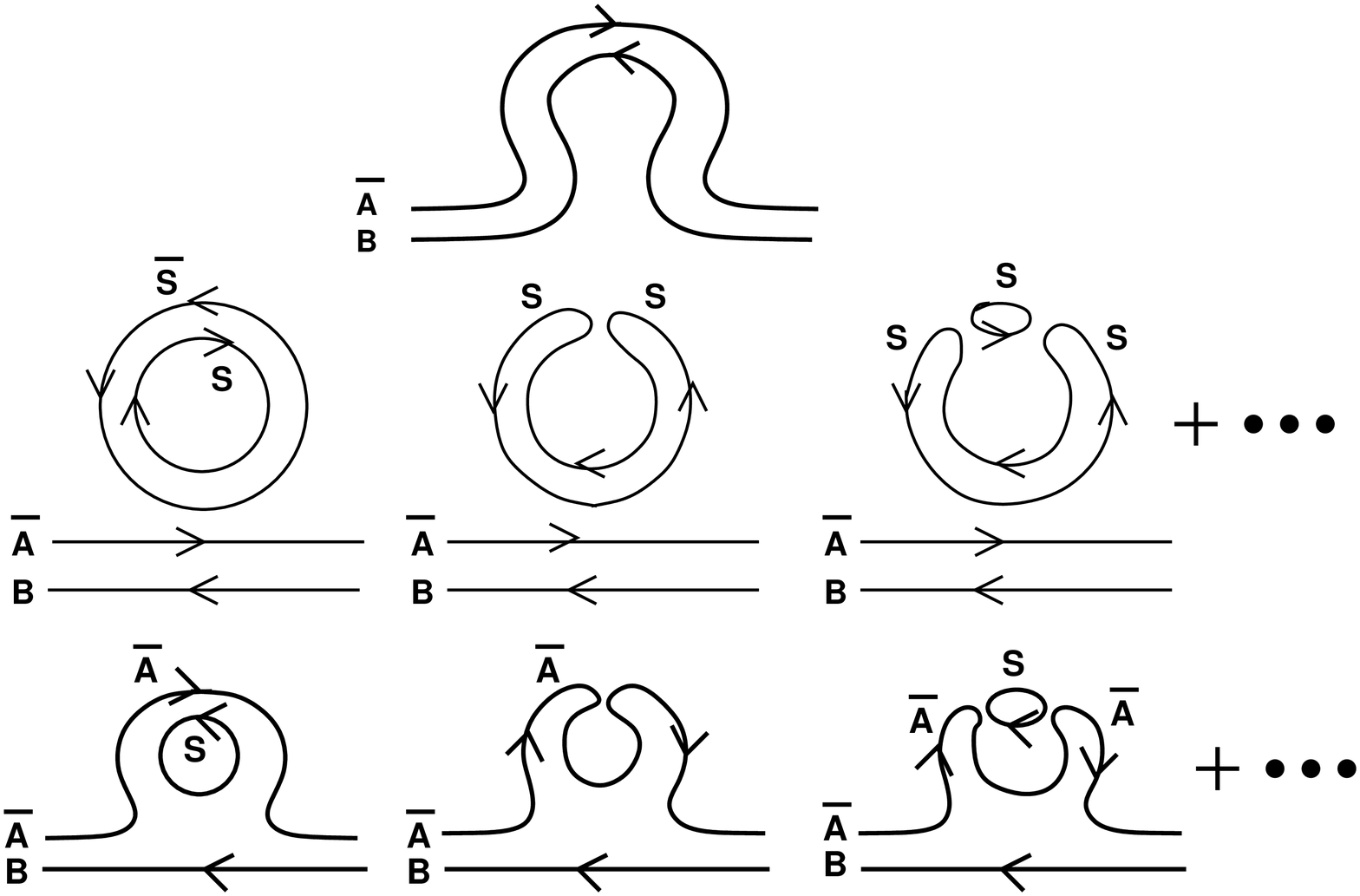}}
	\vspace{-.35in}\caption{Quark line contributions to mass and decay constant renormalization}\vspace{-.25in}
	\label{fig:QL}
\end{CFigure}

\end{document}